# Analytical Expressions and Numerical simulation of single electron spectroscopy

Sheng Wang

*Department of Physics, University of Science and Technology of China, Hefei, 230026, China*

E-mail: ws2010@mail.ustc.edu.cn

**Abstract**

We use the Monte Carlo method to study the two types of devices used in the technique of single electron spectroscopy and get the C-V curve and I-V curve of them. The results compare well to approximate analytical expressions. Furthermore, with great prospects, we may take into account such effects as cotunneling and coupling between quantum dots through the combination of Monte Carlo method and other numerical methods.

## 1. Introduction

The technique of single electron spectroscopy has been widely utilized to study the energy addition spectra of quantum dots. There are two natural ways to carry out such measurements [1].The first is the single electron box model, referred to as single electron capacitance spectroscopy (SECS). SECS is generally used to study the vertical quantum dots. The second is the single electron transistor (SET) model, known as gated transport spectroscopy (GTS). GTS is generally used to study lateral quantum dots.

The SECS experiment is shown in Fig. 1a. The quantum dot will be used as the island of the single electron box. Now the highly charge-sensitive can be incorporated directly on top of a vertical quantum dot as a charge sensor to study single electron addition energy spectra of quantum dots [2]. We get the information of the spectra from the capacitance versus gate voltage curve. A simple geometric scale factor converts the voltage scale into an energy scale to allow a quantitative single electron addition energy spectrum. We will analyze the C-V curve and use Monte Carlo simulation to explain it.

The GTS experiment is shown in Fig. 1b. The quantum dot will be directly used as the island of a weakly biased single electron transistor and measure the source to drain current as a function of gate voltage [3]. As with the SECS experiment, the peak positions reflect the energy required to add each successive electron to the dot. Multiplying the gate voltage scales by a geometric factor converts this scale into an energy scale for the quantum dot.

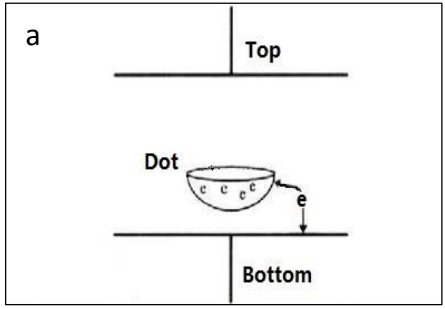 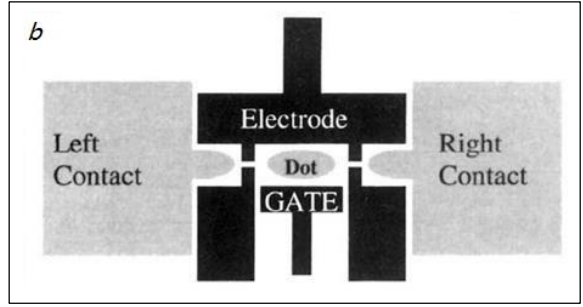

Fig. 1a, Schematic diagram of SECS     Fig. 1b, Schematic diagram of GTS

To simplify our simulation, we will first begin with a single quantum dot based on the orthodox theory, which makes the following major assumptions [1].
1) The thermal energy and quantized energy of the electron can be ignored, so the charging energy can be regarded as the whole energy.
2) The time of electron tunneling through the barrier is assumed to be negligibly small in comparison with other time scales, including the interval between neighboring tunneling events.
3) Coherent quantum processes consisting of several simultaneous events (cotunneling) are ignored. The assumption is valid if the resistance R of all tunnel barriers of the system is much larger than the quantum unit of resistance $R_Q$, Where $R_Q = h/e^2$.

We will later discuss the effects like cotunneling and the interaction between quantum dots in our simulation.

## 2. Monte Carlo method

For many experiments, Quantum Mechanics only predicts the probability of any outcome; therefore, the Monte Carlo method, which is a stochastic technique, is a more suitable method to simulate quantum systems [4].

To elucidate Monte Carlo method, we first have to introduce tunneling rate $\Gamma$ and silence time t. Tunneling rate $\Gamma$ through a single electron practical device depends on the reduction $\Delta W$ of the free energy of the system as a result of this tunneling event. This dependence of the i[th] tunnel junction can be expressed as:

$$\Gamma_i = \frac{\Delta W}{e^2 R_T (1 - \exp(\frac{-\Delta W}{kT}))} \quad (1)$$

where $R_T$ is the tunnel resistance [5].

The silence time $t_i$, the duration to the next tunnel event through the i[th] junction can be expressed as:

$$t_i = -\frac{\ln r_i}{\Gamma_i} \quad (2)$$

where $r_i$ is a random number with uniform distribution and $\Gamma_i$ is the tunnel rate [6].

The simulation algorithm can be briefly summarized as follows:
1. Set the parameters and initialize the whole system.
2. List all possible tunnel events. For each one, compute the charge, the reduction ΔW of the free energy, tunneling rate Γ and silence time t.
3. Select the shortest silence time t and take the corresponding event as the actual tunnel event.
4. Update charges of the system.
5. Repeat step 2 to step 4 to compute the average charge, current, or capacitance across the specified tunnel junctions.
6. Repeat step 2 to step 5 until we have done the computations for all required gate voltages or source to drain voltages.

## 3. Single electron box

The expressions for the charging energy W(n) of the Single-Electron Box (see Fig. 2a) is:

$$W(n) = (ne - Q_{ext})^2 / 2C_\Sigma + \text{constant} \tag{3}$$

where $-ne$ is the island charge, parameter $Q_{ext}$ is defined as $Q_{ext} = CU_0$ ($U_0$ is the gate voltage), $C_\Sigma$ is the total capacitance of the island, including the island-gate capacitance $C_0$ [7].

In our simulation, we will use $Q_{ext}$ as an externally controlled variable, a more convenient way to present the effect of the gate voltage.

At low temperatures, the single electron tunneling will minimize the energy of the system according to (1). In the orthodox theory regime, the reduction ΔW(n) of charging energy can be regarded as the reduction ΔW of the free energy of the system. At the situation where island charge is stable at $-ne$, it will take additional energy to add or subtract one electron from the island.

$$W(n) \leq W(n \pm 1) \tag{4}$$

$$(n - \frac{1}{2})e \leq Q_{ext} \leq (n + \frac{1}{2})e \tag{5}$$

(5) shows that n as a function of $Q_{ext}$ or $U_0$ will be staircase like and there will be sudden changes of island charge whenever $Q_{ext}$ is at (n+1/2e) or gate voltage $U_0$ at (n+1/2e)/$C_0$. Fig. 2b shows the n-$U_0$ curve based on the analytical expressions.

To do the simulation, we have to calculate the reduction ΔW of the free energy, tunneling rate Γ and silence time t of all possible tunnel events.

$$\Delta W_{1,2} = W(n) - W(n \pm 1) \tag{6}$$

Fig. 2c presents the simulation result of the n-$U_0$ curve. The parameters of the single electron box are $R = 2.5 \times 10^5 \Omega$, $C_0 = 2.5 \times 10^{-18}$ F. When the absolute value of the voltage exceeds the critical value $V_c = e/2C_0$, the island charge begins to show the step function behavior. Additionally, the width of each terrace of the n-$U_0$ curve is nearly the same as e/$C_0$. The results compares well to the analytical expressions. We can also understand Fig. 2c in another way. The capacitance as a function of $U_0$ will exhibit peaks at those particular positions due to the sudden changes of island charge as presented in Fig. 2d. The Capacitance-$U_0$ curve is exactly what we measure in SECS experiment. We can get the single electron addition spectra from the positions of the peaks. The peak spacing is determined by the irregular spacing of quantized energy level $\Delta E$ and the charging effects [7]. When $\Delta E \gg e^2/C$, the charging effects will regulate the energy spacing and result in the same spacing between peaks.

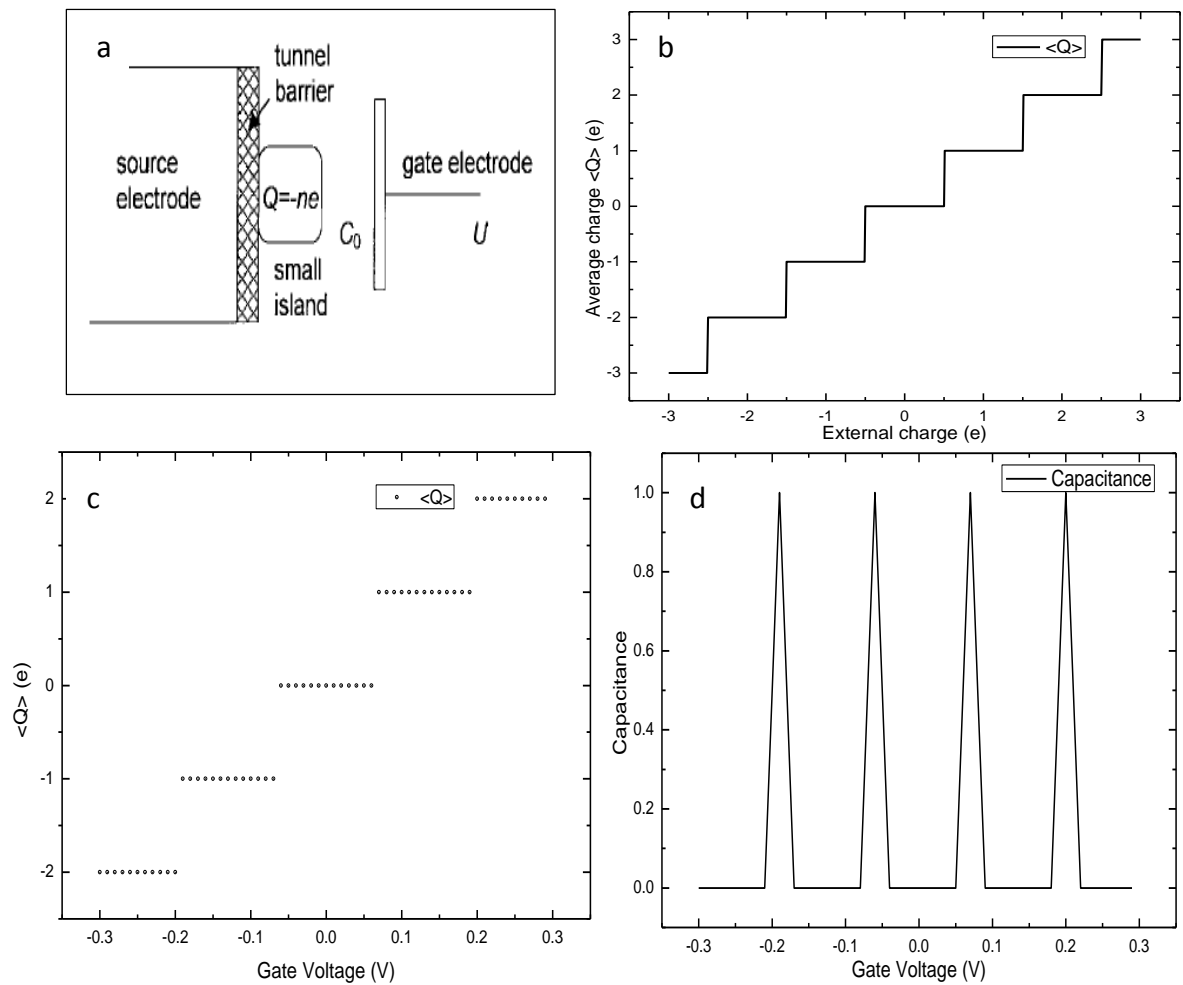

Fig. 2a, Schematic diagram of Single electron box. b, The analytical result of the <Q>-External charge $Q_{ext}$ curve. c, The Monte Carlo simulation result of the <Q>-Gate Voltage curve. d, The capacitance as a function of the Gate Voltage according to the simulation result of c, peaks will appear at those particular positions of Gate Voltage where there are sudden changes of island charge.

### 4. Single electron transistor

As an evident generalization of (3), the charging energy of single electron transistor (See Fig. 3a) can be presented as:

$$W(n_1, n_2) = (ne - Q_{ext})^2 / 2C_\Sigma - eV(n_1 C_2 + n_2 C_1)/C_\Sigma + \text{constant} \quad (7)$$

where $n_1$ and $n_2$ are the number of electrons passed through the tunnel barriers one and two so that $n = n_1 - n_2$, $C_\Sigma$ is the total capacitance of the island, parameter $Q_{ext}$ is still defined as $Q_{ext} = CU_0$ ($U_0$ is the gate voltage) [8].

As with single electron box, at the situation where island charge is stable at $-n_1 e$ and $-n_2 e$, it will take additional energy to add or subtract one electron through either of the tunnel junctions.

$$W(n_1, n_2) \leq W(n_1 \pm 1, n_2 \pm 1) \tag{8}$$

$$(n - \frac{1}{2})e - Q_{ext} \leq VC_2 \leq (n + \frac{1}{2})e - Q_{ext} \tag{9}$$

$$Q_{ext} - (n + \frac{1}{2})e \leq VC_1 \leq Q_{ext} - (n - \frac{1}{2})e \tag{10}$$

For symmetric transistors $C_1 = C_2 = \frac{1}{2}C$ and $n = 0$,

$$\left| V\frac{C}{2} \pm Q_{ext} \right| \leq \frac{e}{2} \tag{11}$$

Fig. 3b presents the threshold voltage as a function of $Q_{ext}$ of symmetric transistors when $n = 0$ according to (11).

For single electron transistor, there are four possible tunnel events through both junction one and junction two.

$$\Delta W_{1,2} = W(n_1, n_2) - W(n_1 \pm 1, n_2) \tag{12}$$

$$\Delta W_{3,4} = W(n_1, n_2) - W(n_1, n_2 \pm 1) \tag{13}$$

Fig. 3c shows the simulation result of the source-drain dc I-V curves of a symmetric transistor for several values of $Q_{ext}$. The parameters we use are:

$$T = 0.3K, C_\Sigma = 7.0 \times 10^{-17} F, C_0 = 2.0 \times 10^{-17} F, R = 2.5 \times 10^5 \Omega \tag{14}$$

From Fig. 3c, there is no Coulomb blockade when $Q_{ext} = e/2$. Yet with lower external charge, the source to drain current will be suppressed at low voltages. When $Q_{ext} = 0$, the threshold voltage is about $e/C$. As $Q_{ext}$ increases from 0 to $e/2$, the threshold voltage will decrease from $e/C$ to 0 in a nearly linear response. The simulation result compares well to the analytical expressions presented in Fig. 3b. Another way to understand the properties of the Fig. 3c is that the linear conductance $dI/dU$, where $U = 0V$, as a function of external charge, i.e., gate voltage will exhibit sharp peaks at those positions where Coulomb blockade is completely suppressed.

Fig. 3d presents the relationship of drain current and gate voltage for a few source-drain voltages. The parameters we use are the same in (14). It is observed that capacitance as a function of gate voltage exhibits peaks at particular positions. When the gate voltage is set precisely at the gate voltage needed for an electron to be added to the dot, the number of electrons on the dot may fluctuate by one. The fluctuation gives rise to a detectable current flowing through the dot and therefore a current peak appears [3]. At low source-drain voltages, there are intervals between peaks where there is no current flow. The interval is the adjustment that gate voltage needed for bringing the next available empty state for electron tunneling [9]. At higher source-drain voltages, Coulomb blockade will be suppressed and the current can flow without suppression. This explains why we apply small voltage difference between the left and right leads in the GTS experiments. In the orthodox theory regime where the charging effects dominate, the spacing between peaks will be the same. When the quantized energy is comparable to charging energy, the information of the quantum dot will be reflected in the different distance between peaks and different height of the peaks.

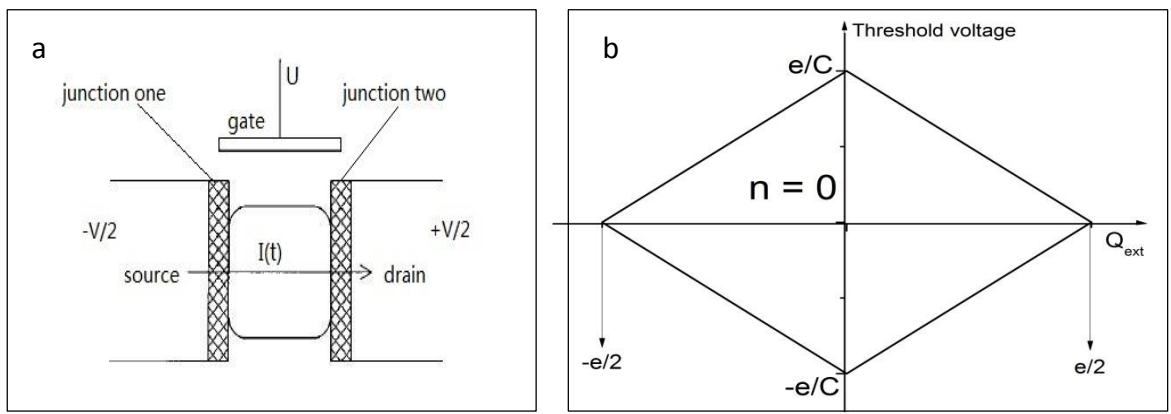

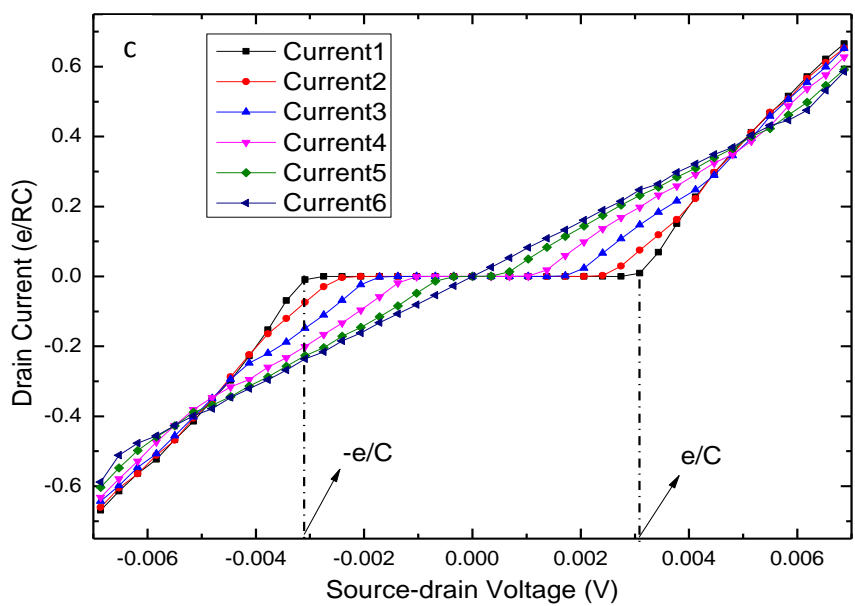

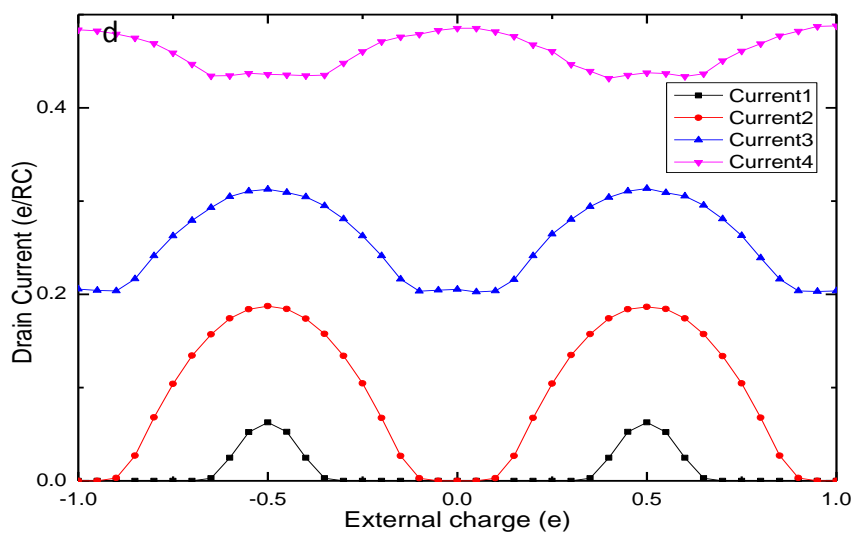

Fig. 3a, Schematic diagram of Single electron transistor. b, The relationship of threshold voltage and external charge $Q_{ext}$. c, The Monte Carlo simulation result of source to drain I-V characteristics of a symmetric SET for a few external charges. d, The Current as a function of the External charge, i.e., gate voltage for a few source-drain voltages.

## 5. Effects beyond the Orthodox Theory regime

Our simulation above is largely based on the orthodox theory, which doesn't take account of some important effects. In the Coulomb blockade regime, where the first order tunnel rate is very low, conduction is dominated by cotunneling processes. Second order co-tunneling looks like a simultaneous tunneling of two electrons through two junctions. It becomes more apparent when the tunnel coupling between the dot and the leads is enhanced. The onset of inelastic cotunneling can be exploited to measure the energy spectrum of a quantum dot with improved resolution [10]. We may explore the properties of this energy spectrum through the combination of Monte Carlo method and Master Equation formalism. Single electron spectroscopy has been used in multiple coupled quantum dot system (also known as "artificial molecules"). When electrons tunnel at appreciable rates between quantum dots in a coupled dot system, the system forms an artificial molecule, where the charge on each dot is no longer quantized and the orthodox theory can no longer be applied [11]. Some models and numerical approaches like boundary-element approach have been used to analyze the system [12]. We can first begin with tunnel coupled double dots, which are controlled by three parameters, the two gate voltages and the inter-dot tunnel conductance. The dot interaction energy and inter-dot tunneling should be taken into consideration. It has been demonstrated that using 0D-states in the second dot of a coupled dot structure to will remove the effects of thermal broadening in the leads and lead to the improvement of the resolution of this spectroscopic technique [13]. Hopefully we may clarify the spectra by applying Monte Carlo method to suitable models.

## 6. Summary and discussion

In this present paper, we have demonstrated the staircase like Q-V curve of single electron box and the Current-Voltage curve of single electron transistor through both analysis and Monte Carlo simulation. The two single electron devices are two natural ways adopted in SECS and GTS, respectively. Though our simulation is largely based on the orthodox theory, the natures of the basic tunneling mechanism of the two devices revealed in this paper give a clear picture of single electron spectroscopy. With great prospects, we can take account of some important effects like cotunneling and coupling between many dots by combining Monte Carlo method and some other numerical methods. Thus, the simulation results may have better agreement with the spectra in single electron spectroscopy experiments, which have been utilized in more complicated system, where some rare effects may play important roles. .